\documentclass[%
nofootinbib,
mathtools,
amsmath,
amssymb,
twocolumn,
prl,
longbibliography,
superscriptaddress]{revtex4-2}

\usepackage[utf8]{inputenc}
\usepackage[T1]{fontenc}
\usepackage{booktabs}
\usepackage{mathrsfs}
\usepackage{graphicx}
\usepackage{dcolumn}
\usepackage{bm}
\usepackage{lipsum}
\usepackage{graphics}
\usepackage{bbm}
\usepackage[dvipsnames]{xcolor}
\usepackage{color}
\usepackage{microtype}
\usepackage{IEEEtrantools}
\usepackage{mathrsfs}
\usepackage{amsthm}
\usepackage{enumitem}
\usepackage[normalem]{ulem}
\usepackage[pdftex,breaklinks,colorlinks,
linkcolor=Blue,
citecolor=teal,
anchorcolor=red,
urlcolor=cyan]{hyperref}
\usepackage{orcidlink}
\usepackage{mathtools}
\graphicspath{{Figures/}}

\newcommand{\e}{\varepsilon}

\def\l{\ell}
\def\m{m}
\def\n{n}

\def\L{\ell}
\def\M{m}

\def\lOne{\ell_1}
\def\mOne{m_1}
\def\nOne{n_1}

\def\lTwo{\ell_2}
\def\mTwo{m_2}

\def\nTwo{n_2}

\begin{document}

\title{Dissection of a merger-ringdown waveform in the small-mass-ratio limit}

\author{Patrick Bourg\,\orcidlink{0000-0003-0015-0861}}
\email{patrick.bourg@ru.nl}
\affiliation{Institute for Mathematics, Astrophysics and Particle Physics, Radboud University, Heyendaalseweg 135, 6525 AJ Nijmegen, The Netherlands}
\author{Lorenzo K\"uchler\,\orcidlink{0000-0002-6609-1684}}
\email{l.m.kuchler@soton.ac.uk}
\author{Benjamin Leather\,\orcidlink{0000-0001-6186-7271}}
\email{bjl1u25@soton.ac.uk}
\author{Kevin Cunningham\,\orcidlink{0000-0002-1094-2267}}
\author{Jack Lewis\,\orcidlink{0000-0001-8345-3176}}
\author{Adam Pound\,\orcidlink{0000-0001-9446-0638}}
\email{a.pound@soton.ac.uk}
\affiliation{School of Mathematical Sciences and STAG Research Centre, University of Southampton, Southampton, United Kingdom, SO17 1BJ}
\author{B\'eatrice Bonga\,\orcidlink{0000-0002-5808-9517}}
\email{bbonga@science.ru.nl}
\affiliation{Institute for Mathematics, Astrophysics and Particle Physics, Radboud University, Heyendaalseweg 135, 6525 AJ Nijmegen, The Netherlands}
\author{Marc Casals\,\orcidlink{0000-0002-8914-4072}}\email{marc.casals@uni-leipzig.de}
\affiliation{Institut f\"ur Theoretische Physik, Universit\"at Leipzig,\\ Br\"uderstra{\ss}e 16, 04103 Leipzig, Germany}
\affiliation{Scoil na Matamaitice agus na Staitistic\'i, An Col\'aiste Ollscoile, Baile \'Atha Cliath,  D04 N2E5, Ireland}
\affiliation{Centro Brasileiro de Pesquisas F\'isicas (CBPF), Rio de Janeiro, CEP 22290-180, Brazil}
\author{Rodrigo Panosso Macedo\,\orcidlink{0000-0003-2942-5080}}\email{rodrigo.macedo@nbi.ku.dk}
\affiliation{Center of Gravity, Niels Bohr Institute, Blegdamsvej 17, 2100 Copenhagen, Denmark}

\date{\today}

\begin{abstract}
Work over the past two decades has unveiled the rich phenomenology of black hole binary mergers and subsequent ringdowns, involving a tapestry of quasinormal modes (QNMs), nonlinearities, tails, transients, and secular effects including gravitational memory. Here we develop a framework for analyzing nonlinear merger-ringdown features in the small-mass-ratio limit, where individual effects can be cleanly isolated. Specializing to the case of a quasicircular, nonspinning black hole binary, we find the waveform sharply divides into a pre-merger extended inspiral phase, a merger phase lasting roughly half a cycle, and a post-merger ringdown dominated by QNMs. We show quadratic QNMs dominate over linear overtones in the ringdown phase for comparable-to-intermediate mass ratios, and we highlight nonlinear effects of gravitational-wave memory, including cubic wave-zone phenomena analogous to horizon absorption effects.
\end{abstract}

\maketitle

\emph{Introduction.}
Binary black hole (BH) mergers produce gravitational waves whose full signal---from the early inspiral through the cataclysmic merger to the late-time ringdown---encodes the properties of the progenitor and remnant BHs. But unlocking these properties, and performing precision tests of BHs and general relativity (GR), is only possible through accurate waveform modeling~\cite{LISA:2022kgy,LISAConsortiumWaveformWorkingGroup:2023arg,Dhani:2024jja,Chandramouli:2024vhw,ET:2025xjr}. Understanding the merger-ringdown phase has proved particularly vital and challenging in this regard, as it is crucial for the BH spectroscopy program~\cite{Berti:2025hly,LIGOScientific:2025wao} while generally being the most resistant to accurate semi-analytical and phenomenological modeling~\cite{Varma:2018mmi,Albertini:2022rfe,Thompson:2023ase,Dhani:2024jja,Foo:2024exr,Mahesh:2025oaf,Siegel:2025xgb,Crescimbeni:2025ytx,Mahapatra:2026wsp,Gamboa:2026jht}.

Since the earliest incarnation of an inspiral-merger-ringdown (IMR) waveform model~\cite{Buonanno:2000ef}, an important touchstone  has been the small-mass-ratio limit, where one BH (the secondary) is much smaller than the other (the primary). This limit has been used both to inform IMR models' treatment of the merger~\cite{Buonanno:2000ef,Damour:2007xr,Bernuzzi:2010xj,Taracchini:2012ig,Taracchini:2013rva,Taracchini:2014zpa,McWilliams:2018ztb,Rifat:2019ltp,Islam:2022laz,Albanesi:2023bgi,Albanesi:2026qtx,Faggioli:2026alx,Nishimura:2026nse} and to investigate the merger-ringdown's underlying physical processes in a clean perturbative setting~\cite{Mino:2008at,Hadar:2009ip,Zimmerman:2011dx,Folacci:2018cic,Hughes:2019zmt,Oshita:2023cjz,DeAmicis:2024eoy,Becker:2025zzw,Kuchler:2025hwx,DeAmicis:2025xuh,Oshita:2025qmn,Roy:2025kra,DellaRocca:2025zbe,DeAmicis:2026wqd,Ma:2026qbq,Su:2026gmp}. However, such studies have historically been restricted to linear perturbation theory, in which the secondary plunges into the primary, generates a linear perturbation in the process, and leaves the primary ringing down with characteristic quasinormal mode (QNM) frequencies~\cite{Taracchini:2014zpa}. This linear approximation is fundamentally limited both in accuracy and in the physics it can describe---a critical limitation given the emerging recognition of the importance of \emph{nonlinear} effects in the merger-ringdown, from quadratic QNMs (QQNMs)~\cite{London:2014cma, Mitman:2022qdl, Cheung:2022rbm, Baibhav:2023clw, Cheung:2023vki, Khera:2023oyf, Yi:2024elj, Giesler:2024hcr, Lagos:2024ekd,Wang:2026rev} to nonlinear memory~\cite{Pollney:2010hs,MaganaZertuche:2021syq,Gasparotto:2023fcg,Goncharov:2023woe,Rossello-Sastre:2024zlr,Inchauspe:2024ibs}.

Nonlinear BH perturbation theory~\cite{Gleiser:1995gx, Gleiser:1998rw, Campanelli:1998jv, Brizuela:2006ne, Brizuela:2009qd, Green:2019nam, Spiers:2023cip, Spiers:2023mor} offers a way to overcome this limitation. Recent advances have spurred numerous calculations at second perturbative order in the ringdown, but disconnected from any preceding merger~\cite{Nakano:2007cj, Ioka:2007ak, Okuzumi:2008ej, Loutrel:2020wbw, Ripley:2020xby, Lagos:2022otp, Redondo-Yuste:2023seq, Kehagias:2023ctr, Ma:2024qcv, Bourg:2024jme, Zhu:2024rej, Bucciotti:2024zyp, Bucciotti:2024jrv, Khera:2024bjs, Kehagias:2024sgh, Bourg:2025lpd, Bucciotti:2025rxa}. In parallel, motivated by the need for high-accuracy models of extreme-mass-ratio inspirals~\cite{LISA:2022yao}, self-force theory~\cite{Barack:2018yvs} has extended inspiral models to second order in the small mass ratio, but stopping short of the merger~\cite{Rosenthal:2006iy, Detweiler:2011tt, Pound:2012nt, Gralla:2012db, Pound:2012dk, Pound:2014xva, Pound:2017psq, Pound:2019lzj, Miller:2020bft, Upton:2021oxf, Warburton:2021kwk, Wardell:2021fyy, Bonetto:2021exn, Miller:2023ers, Cunningham:2024dog, Upton:2025bja, Mathews:2025txc, Honet:2025lmk}. Most recently, two of us have developed a self-force formalism for complete, first-principles IMR waveform models to any perturbative order in the mass ratio~\cite{Kuchler:2024esj,Kuchler:2025hwx,Honet:2025dho,Roy:2025kra}, offering the prospect of end-to-end calculations of nonlinear perturbative effects from the inspiral through to the final ringdown.

In this letter, we tie these threads together. While a complete second-order merger-ringdown calculation remains a severe technical challenge, we show that a careful dissection of the first-order (in mass ratio) solution provides a controlled, systematic route to compute second-order (and higher) effects in a modular way. 

\begin{figure}[t] \centering
\includegraphics[width=\columnwidth]{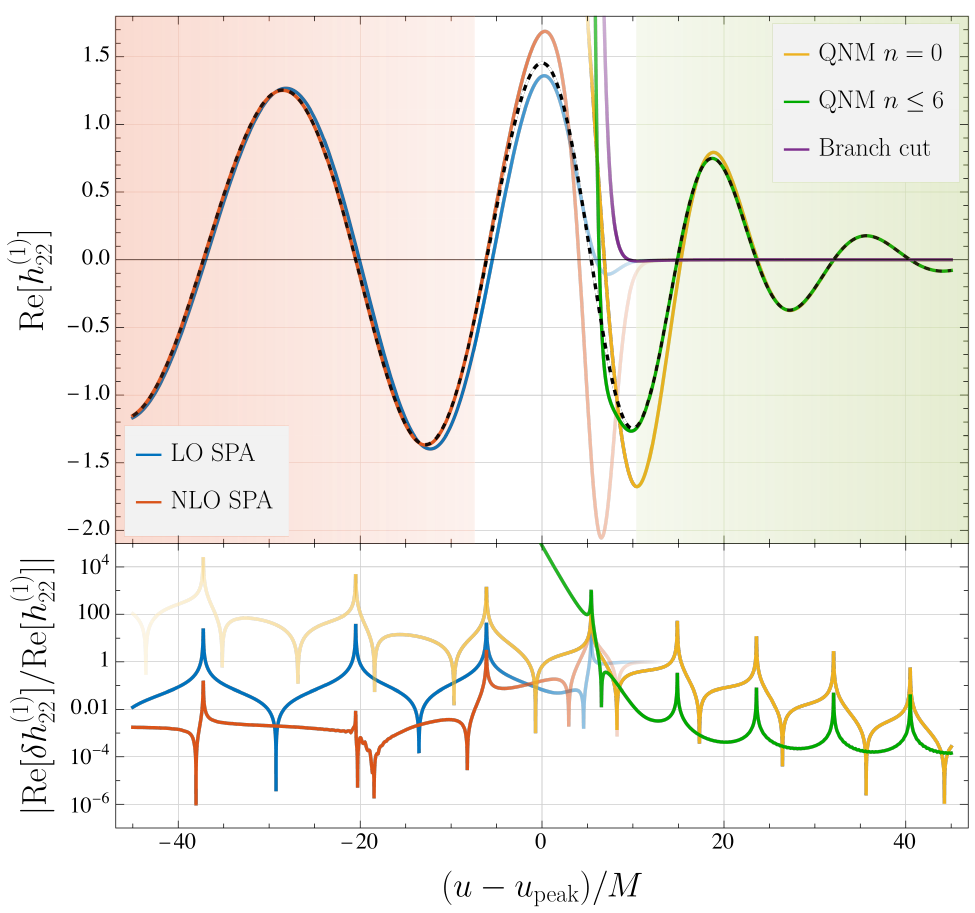}
    \caption{Top: 
    the dominant first-order waveform mode, $h_{22}^{(1)}$. The full plunge waveform (black dashed curve) is obtained from Eq.~\eqref{eq:h1lm}. 
    At early times (pink shaded region), it is accurately captured by a SPA, with NLO terms improving the result. At late times (green shaded region), it is well described by the sum over QNMs in Eq.~\eqref{eq:linearQNMsum}, with additional overtones improving the agreement and with negligible contribution from the branch cut in Eq.~\eqref{eq:G_BC_integral}. Neither approximation reaches the peak of $|h^{(1)}_{22}|$ at $u=u_\text{peak}$. The shaded regions indicate the regimes in which the NLO SPA and the $n\leq6$ QNM sum have relative errors $\leq10^{-2}$ with respect to $h^{(1)}_{22}$ ($u-u_\text{peak}\lesssim-7.40M$ and $u-u_\text{peak}\gtrsim10.36M$, respectively). QNM and branch-cut curves remain finite at all times but are truncated once they deviate significantly from the full waveform. Bottom: relative error of the SPA and QNM sum with respect to $h^{(1)}_{22}$.
    } 
    \label{fig:Waveform_firstorder_l2m2}
\end{figure}

Restricting to quasicircular, nonspinning binaries, we divide the first-order merger-ringdown waveform into three stages, as displayed in Fig.~\ref{fig:Waveform_firstorder_l2m2}: an extended inspiral, when the waveform is dominated by direct emission from the plunging secondary; the final ringdown, which is dominated by QNMs; 
and the brief merger in between. More recent, alternative decompositions could be adopted~\cite{DeAmicis:2025xuh,DeAmicis:2026wqd,Arnaudo:2025uos,Kuntz:2025gdq, Rosato:2026moe, Ma:2026hcb, Ma:2026qbq, Su:2026fvj, Su:2026gmp, Sun:2026mto}, but we opt for the most traditional division as a demonstration. Our analysis additionally serves to sharpen the traditional split, clearly delineating the boundaries of each regime and isolating the relative contributions of different effects within them.

\begin{figure}[t] \centering \includegraphics[width=.85\columnwidth]{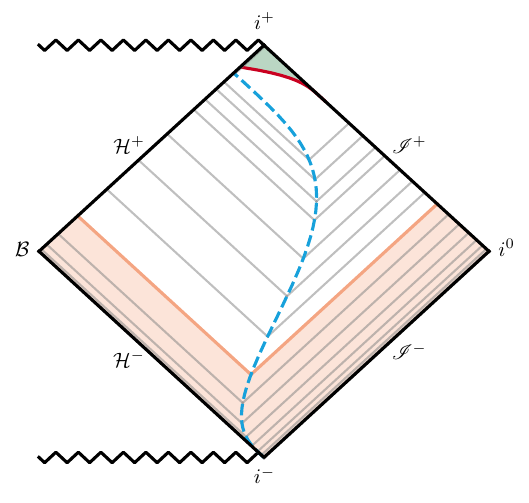}
    \caption{
    Penrose diagram with our geodesic plunge trajectory (blue dashed curve) and null slices (gray lines). Green shading indicates the QNM regime, defined by the linear QNM sum achieving a residual error strictly $\leq 10^{-2}$ relative to the full linear waveform, and bounded by the hyperboloidal slice (red) used as the initial time for our QQNM  calculations. Pink shading indicates the extended-inspiral regime of validity of the SPA, which spans the orbital radii $3.90\leq r_p/M\leq6$. To highlight the three regimes we use a tailored compactification described in the End Matter.
}
    \label{fig:Penrose_diagram}
\end{figure}

Given our first-order dissection, we focus on two second-order (and higher) effects that can be cleanly computed without a full second-order calculation: nonlinear memory effects and QQNMs. From the linear amplitudes of the QNMs, we compute the quadratic QNM couplings using our method from~\cite{Bourg:2024jme,Bourg:2025lpd}, finding that QQNMs dominate over linear overtones for a large range of mass ratios. We also find that the dominant nonlinear memory mode is of similar magnitude as the dominant QQNM, and we highlight a new cubic ``memory of memory'' effect, in which secular nonlinear memory excites new modes, analogous to absorption-induced excitations~\cite{Sberna:2021eui,May:2024rrg}. 

This marks the first time nonlinear effects have been calculated from first principles in a realistic merger scenario, in a pristine setting where such effects are precisely defined. The results have direct implications for gravitational-wave data analysis. Recent studies already claim the observation of anywhere from one to six (!) QQNMs in GW250114~\cite{Yang:2025ror,Wang:2026rev}, underscoring the urgency of precise theoretical predictions. By providing a systematic, efficient framework for isolating and quantifying nonlinear effects in the waveform, our analysis offers concrete guidance for merger-ringdown modeling---and consequently for tests of the Kerr metric and GR.

\emph{Linear waveform from a plunging particle.} We first calculate the complete linear-order merger-ringdown waveform (black dashed curve in Fig.~\ref{fig:Waveform_firstorder_l2m2}), prior to its dissection. We consider mergers at the end of quasicircular inspirals, in which the secondary, treated as a particle of mass $m_p$, spirals into a Schwarzschild BH of mass $M$. Expanding in powers of the mass ratio, $\e\coloneqq m_p/M$, we build the spacetime metric $g_{\mu\nu}+\e h^{(1)}_{\mu\nu} + \e^2 h^{(2)}_{\mu\nu}+\ldots$, where $g_{\mu\nu}$ is the Schwarzschild background, and any corrections to the primary are incorporated into the perturbations. At leading order, the merger-ringdown waveform is then generated by the particle slowly plunging from the innermost stable circular orbit (ISCO) along a geodesic $x^\alpha_p$ of the Schwarzschild spacetime
~\cite{Kuchler:2025hwx}, depicted in Fig.~\ref{fig:Penrose_diagram}. 

\begin{figure} \centering
    \includegraphics[width=.875\columnwidth]{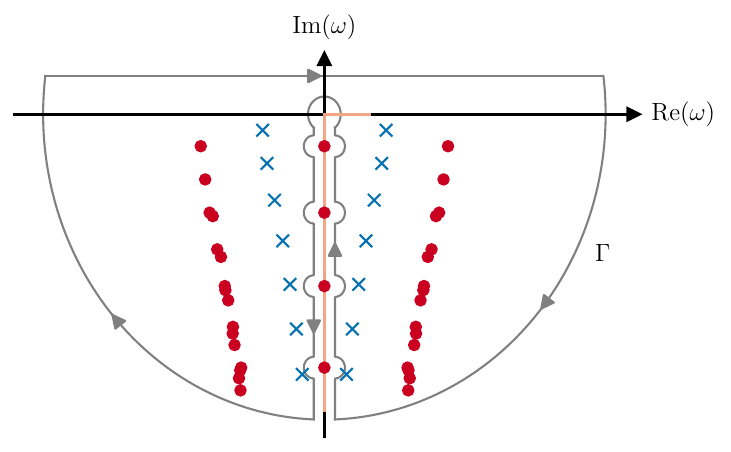}
    \caption{
    Integration contour (grey) for the various components of our first- and second-order waveform.
    Blue crosses mark the poles corresponding to the (linear) QNMs for $\ell = 2$, while the red circles denote the QQNM poles for $(\L, \M) = (4, 4)$ that arise from couplings of the linear QNMs.
    The orange line along the real axis indicates the range covered by the SPA, while the orange line along the imaginary axis indicates the range used for the branch-cut contribution, both for $(\ell,\m)=(2,2)$.
}
    \label{fig:integration_contour}
\end{figure}

Working in Schwarzschild coordinates $x^\alpha=(t,r,\theta,\phi)$, 
we foliate the BH's exterior with slices of constant time $s=t-\kappa(r)$. Choosing these slices to be null (grey lines in Fig.~\ref{fig:Penrose_diagram}), we link each point on the particle's trajectory and each advanced time $v$ on the BH future horizon ($\mathscr{H}^+$) to a retarded time $u$ at future null infinity ($\mathscr{I}^+$). The waveform strain, $h\coloneqq\lim_{r\to\infty}\frac{r}{M}(\e h^{(1)}_{\bar m\bar m}+ \e^2 h^{(2)}_{\bar m\bar m}+\ldots)$ with $\bar m^\mu=\frac{1}{\sqrt{2}r}(0,0,1,-i\csc\theta)$, can then be written in terms of dynamical properties of the two objects. Decomposing $h$ into spin-weight $-2$ spherical harmonics~\cite{goldberg1967spin}, $h=\sum_{\ell\geq2}\sum_{|m|\leq\ell} h_{\ell m} \; {}_{-2}Y_{\ell m}$, yields its $(\ell,m)$ modes.

To compute the linear waveform, $h^{(1)}_{\ell m}$, we solve the Regge-Wheeler-Zerilli (RWZ) equations~\cite{PhysRev.108.1063,PhysRevD.2.2141,Cunningham:1978zfa,Moncrief:1974am,Cunningham:1979px} sourced by the plunging particle. Using the standard frequency-domain (FD) Green-function method~\cite{Pound:2021qin,Leaver:1986gd}, we obtain the FD strain of the form $\hat h^{(1)}_{\ell m}(\omega)=\frac{\sqrt{D_\ell}}{2}\lim_{r\to\infty}\int \frac{dr'}{f(r')} G_\ell(\omega,r,r')S_{\ell m}(\omega,r')$ in terms of the FD point-particle source $S_{\ell m}$, where $f(r)\coloneqq 1-2M/r$ and $D_\ell\coloneqq(\ell-1)\ell(\ell+1)(\ell+2)$; the time-domain strain is then $h^{(1)}_{\ell m}(u) = \frac{1}{2\pi}\int_{-\infty}^\infty d\omega e^{-i\omega u }\hat h^{(1)}_{\ell m}$. The retarded Green function $G_\ell$ is constructed from two independent homogeneous solutions to the FD RWZ equations, the ``in''  and ``up'' solutions, $R^\text{e/o in}_{\ell}$ (regular at $\mathscr{H}^+$) and $R^\text{e/o up}_{\ell}$ (regular at $\mathscr{I}^+$). Here, $\text{e/o}$ denotes the even- and odd-parity sectors. 
The strain then reduces to~\cite{Kuchler:2025hwx} 
\begin{equation}\label{eq:h1lm}
    h_{\ell m}^{(1)}(u) = \frac{\sqrt{D_\ell}}{4\pi}\int_{-\infty}^{+\infty} d\omega \, e^{-i\omega u} \frac{C_{\ell m}^\text{e}(\omega) - i C_{\ell m}^\text{o}(\omega)}{2i\omega A^\text{in,inc}_\ell(\omega)},
\end{equation}
where $A^\text{in,inc}_\ell$ is the incidence coefficient of the in solution (which does not depend on parity), and $C_{\ell m}^\text{e/o}$ is given by an integral along the particle's trajectory,
\begin{equation}\label{eq:Clm}
    C_{\ell m}^\text{e/o}(\omega) = -\int_{2M}^{6M}\frac{dr'}{f(r')} R^\text{e/o in}_{\ell}(\omega,r') S^\text{e/o}_{\ell m}(\omega, r').
\end{equation}
Since the plunge trajectory approaches $r_p=6M$ in the infinite past, the integration extends to the ISCO radius only. In practice we evaluate the integral~\eqref{eq:h1lm} over the range $-4\leq M\omega\leq 4$. 
Such leading-order waveforms were computed in Ref.~\cite{Kuchler:2025hwx} for all modes with $\ell=2,\dots,12$ and $m\neq0$; here, we have also computed the $m=0$ modes for $\ell=2,3,4$. Figure~\ref{fig:Waveform_firstorder_l2m2} displays the $(\ell,m)=(2,2)$ mode.

We now partition the plunge waveform into the three distinct regimes in Fig.~\ref{fig:Waveform_firstorder_l2m2}, extending and sharpening the division we previously highlighted in Ref.~\cite{Kuchler:2025hwx}.

\emph{Extended inspiral.} At early times, the particle's orbital radius $r_p$ changes slowly relative to its orbital phase $\phi_p$, and the waveform frequency closely tracks the orbital frequency $\Omega=d\phi_p/dt$, extending the hallmarks of the inspiral. We can exploit this using a stationary phase approximation (SPA). The waveform $h^{(1)}$ obtained from Eqs.~\eqref{eq:h1lm} and \eqref{eq:Clm} can be written as a double integral over the frequency $\omega$ and radius $r'$, which can be evaluated using a two-dimensional SPA, with the stationary point $r'=r_p$ and $\omega=m\Omega(r_p)$. In this work, we extend  Ref.~\cite{Kuchler:2025hwx} by including next-to-leading-order (NLO) terms in the SPA, given in the End Matter. 

Figure~\ref{fig:Waveform_firstorder_l2m2} shows the SPA agrees well with the full linear waveform and significantly improves at NLO. However, the SPA breaks down before the waveform's peak and ringdown onset, when the waveform is no longer dominated by direct emission from the particle. This can partly be understood from Fig.~\ref{fig:integration_contour}: the SPA picks out frequencies $\omega=m\Omega$, and $\Omega$ peaks at the light ring $\Omega(3M)=1/(3\sqrt{6}M)$; for $(\ell,m)=(2,2)$ this gives $\omega\approx0.27/M$, while in the ringdown the waveform is dominated by the fundamental QNM frequency, at ${\rm Re}(\omega)\approx0.37/M$.

\emph{Ringdown.} Rather than directly evaluating the integral~\eqref{eq:h1lm} along the real frequency axis, one may follow Leaver~\cite{Leaver:1986gd} by closing the contour in the complex $\omega$ plane (Fig.~\ref{fig:integration_contour}). The Green function has poles at the QNM frequencies and a branch cut along the negative imaginary axis. Applying the residue theorem, we can equate the integral along the real axis to a sum of residues at the poles, an integral around the branch cut, and an integral along the high-frequency arcs~\cite{Stucker-PhD-2026}.

The QNM frequencies $\omega_{\ell n,\pm}$ separate into regular (${\rm Re}(\omega_{\ell n,+})>0$) and mirror modes ($\omega_{\ell n,-}=-\left(\omega_{\ell n,+}\right)^*$), where $n=0$ denotes the fundamental mode and $n\ge1$ the overtones. Summing over the residues at the QNM frequencies, we obtain
\begin{equation}\label{eq:linearQNMsum}
    h_{\ell m}^{(1)\text{QNM}}(u) = \sum_{n=0}^\infty h^{(1)}_{\ell m n,\pm}\, e^{-i\omega_{\ell n,\pm} u},
\end{equation}
with the excitation coefficients~\cite{Berti:2006wq,Zhang:2013ksa,Oshita:2021iyn}
\begin{equation}
    h^{(1)}_{\ell m n,\pm}\coloneqq \frac{\sqrt{D_\ell}}{4\omega} \frac{C^\text{e}_{\ell m}(\omega) - i C^\text{o}_{\ell m}(\omega)}{dA^\text{in,inc}_\ell/d\omega}\biggr|_{\omega=\omega_{\ell n,\pm}}.
\end{equation} 
We compute these coefficients for $\ell=2,\dots,7$, $|m|\leq\ell$, and overtones up to $n=6$ (for $\ell=2)$ or $n=3$ (for $\ell>2$), extending Ref.~\cite{Kuchler:2025hwx}'s calculations. 
Figure~\ref{fig:Waveform_firstorder_l2m2} shows that the QNM sum well approximates the full waveform at late times, and adding overtones improves the accuracy. However, like in the SPA, additional terms do \emph{not} extend the sum's validity toward the waveform peak.

To complete the waveform, we should add the contribution of the branch cut and arcs in Fig.~\ref{fig:integration_contour}. Here we confine our attention to the branch cut, 
which is primarily associated with a power-law tail at late times~\cite{Price:1971fb,Price:1972pw} but is little studied at earlier times~\cite{Casals:2015nja,Casals:2013mpa}. Its contribution can be evaluated by integrating the ($r$-independent part of the) Green function's jump across the branch cut~\cite{Casals:2015nja}:
\begin{equation} \label{eq:G_BC_integral}
h^{(1)\rm BC}_{\ell m}(u) = 
\frac{\sqrt{D_\ell}}{4\pi}
\!\int\limits_{0}^{-i\infty }\!\!\! d\omega\ 
e^{-i \omega u}
[C^{\text{e}}_{\ell m}(\omega) - i C^{\text{o}}_{\ell m}(\omega)]\delta G_{\ell}(\omega).
\end{equation}
The jump $\delta G_{\ell}(\omega)$ is obtained from the general finite-radius expression given in~\cite{Leaver:1986gd,Casals:2015nja}, leading to 
\begin{align} 
\label{eq:bc_discontinuity}
&
\delta G_{\ell}(\omega)
= \frac{q(\omega)}{2\omega^{*}\bigl|A^{\rm in, inc}_{\ell}\bigr|^2}
A^{{\rm in}, {\rm ref}}_{\ell} \, ,
\end{align}
where $q(\omega)$ is the ``branch-cut strength''~\cite{Casals:2015nja}. We evaluate the integral~\eqref{eq:G_BC_integral} down to $M\omega = -1.5i$, far enough for the integrand to have decayed by several orders of magnitude.

\begin{figure}[t] \centering
\includegraphics[width=\columnwidth]{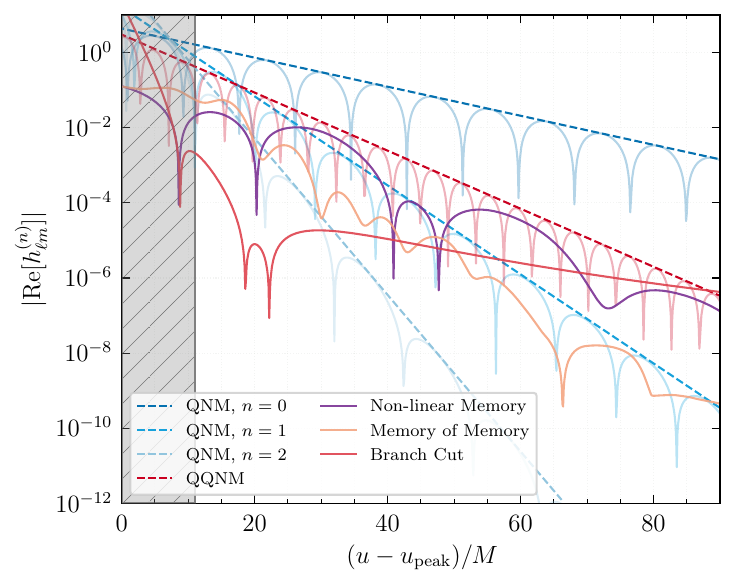}
    \caption{
    Individual contributions to the ringdown waveform.
    We show the $(\ell, \m)$ mode with the largest amplitude for each contribution: $(\L, \M) = (4,4)$ for the QQNM, $(\ell, \m) = (2,2)$ for the QNMs and branch cut, and $(\ell, \m) = (2,0)$ for the nonlinear memory and memory of memory.}
    \label{fig:Waveform_dissection}
\end{figure}

In Figs.~\ref{fig:Waveform_firstorder_l2m2} and \ref{fig:Waveform_dissection}, we see the branch cut contributes negligibly to the waveform at times when the QNM sum is well behaved (though we expect it to be more significant for eccentric binaries~\cite{DeAmicis:2024eoy}). In the earlier, merger regime, both the branch cut and QNM contributions rapidly grow large, and combining them does not approximate the full linear waveform. The residue theorem guarantees that we will recover the full waveform by including the high-frequency arc (for finite arc size), but since all three contributions are exponentially large, their cancellation will be delicate. We hence conclude that the Leaver decomposition only becomes sensible significantly after the waveform's peak, echoing Refs.~\cite{DeAmicis:2025xuh,Arnaudo:2025uos,Kuntz:2025gdq,DeAmicis:2026wqd}. 


\emph{Second order.} At order $\e^2$, $h^{(2)}_{\mu\nu}$ is sourced by quadratic combinations of $h^{(1)}_{\mu\nu}$, along with corrections to the particle's stress-energy tensor~\cite{Kuchler:2025hwx}. Solving for $h^{(2)}_{\mu\nu}$ across the whole of the spacetime is exceedingly challenging due to the extreme singularity at the particle~\cite{Upton:2025bja}, which is exacerbated in the highly relativistic plunge. But in Fig.~\ref{fig:Penrose_diagram} we can identify regions of spacetime where pieces of $h^{(2)}_{\mu\nu}$ can be computed without solving the field equations globally. In the ringdown regime (shaded green in Fig.~\ref{fig:Penrose_diagram}), $h^{(1)}_{\mu\nu}$ is dominated by QNMs (together with nonradiative, $\ell<2$ modes encoding the particle's contribution to the remnant BH's mass and spin), the amplitudes of which we know precisely; this allows us to compute the QQNMs following Ref.~\cite{Bourg:2025lpd}. Along $\mathscr{I}^+$ itself, we can compute memory effects directly from the waveform modes $h^{(1)}_{\ell m}$, without knowledge of the metric in the interior. In principle, we can also apply an SPA to $h^{(1)}_{\mu\nu}$ and $h^{(2)}_{\mu\nu}$ throughout the early spacetime (shaded pink in Fig.~\ref{fig:Penrose_diagram}), as in Ref.~\cite{Lewis:2025ydo}'s treatment of the inspiral, and obtain $h^{(2)}_{\mu\nu}$ in that region using the same methods as for the inspiral~\cite{Pound:2019lzj, Miller:2020bft, Upton:2021oxf, Warburton:2021kwk, Wardell:2021fyy, Bonetto:2021exn, Miller:2023ers, Cunningham:2024dog, Upton:2025bja}, though we leave that for future work.

\emph{Quadratic QNMs.} 
Any two QNMs, with frequencies $\omega_{\bm{n}_1}$ and $\omega_{\bm{n}_2}$, will create a source with time dependence $\exp[-i(\omega_{\bm{n}_1} + \omega_{\bm{n}_2})s]$, generating a QQNM with frequency $\omega_{\bm{n}_1\bm{n}_2}  = \omega_{\bm{n}_1} + \omega_{\bm{n}_2}$; here we define $\bm{n}_i = (\l_i,\m_i,\n_i,\pm)$ and take $s$ to be any hyperboloidal time~\cite{Zenginoglu:2011jz,PanossoMacedo:2023qzp,PanossoMacedo:2024nkw}, not necessarily the null time in Fig.~\ref{fig:Penrose_diagram}. In Refs.~\cite{Bourg:2024jme,Bourg:2025lpd} we provided a method of computing the QQNM amplitudes in any vacuum region to the future of a constant-$s$ slice, given the amplitudes of the parent QNMs. We take that initial slice to be the red curve in Fig.~\ref{fig:Penrose_diagram}, connecting to $\mathscr{I}
^+$ at the retarded time when the QNM sum becomes accurate in Fig.~\ref{fig:Waveform_firstorder_l2m2}. 
The total QQNM contribution to an $(\L,\M)$ mode of the strain $h^{(2)}$ can be decomposed as
\begin{align}
     h^{(2)\rm QQNM}_{\L\M}(u) &= \sum_{\bm{n}_1,\bm{n}_2} h^{(2,\bm{n}_1\bm{n_2})}_{\L\M} \, e^{-i\omega_{\bm{n}_1\bm{n}_2}u} \label{eqn:Upsilon_TD}
\end{align}
with amplitudes $h^{(2,\bm{n}_1\bm{n_2})}_{\L\M}$ directly related to the parent amplitudes $h^{(1)}_{\l\m\n,\pm}$ in Eq.~\eqref{eq:linearQNMsum}~\cite{Bourg:2025lpd}:
\begin{align}
    &h_{{\L, \mOne+\mTwo}}^{(2,\bm{n}_1\bm{n_2})} =  a_{\L, \mOne+\mTwo}  h_{\lOne \mOne \nOne,+}^{(1)}  h_{\lTwo \mTwo \nTwo,+}^{(1)}  \nonumber\\*
    &\ \quad\qquad\qquad +  b_{\L, \mOne+\mTwo} h_{\lOne \mOne \nOne,+}^{(1)} \bigl( h_{\lTwo, -\mTwo, \nTwo,-}^{(1)}\bigr)^\star  \\*
    &\ \quad\qquad\qquad +  c_{\L, \mOne+\mTwo} \bigl( h_{\lOne, -\mOne, \nOne,-}^{(1)}\bigr)^\star h_{\lTwo, \mTwo, \nTwo,+}^{(1)}\,,\nonumber
\end{align}
observing that $h_{{\L\M}}^{(2,\bm{n}_1\bm{n_2})}$ vanishes unless $\M=m_1+m_2$. The complex constants $a_{\L\M},b_{\L\M},c_{\L\M}$ only depend on the background metric; see Table~I of Ref.~\cite{Bourg:2025lpd}. We make these constants publicly available for all possible first-order mode couplings with $2 \leq \l \leq 5$, $\n < 3$, and all nontrivial $\m$~\cite{QQNMZenodo}.
We focus on QQNMs formed by the coupling of two regular or two mirror modes, as their linear amplitudes dominate over mixed regular-mirror couplings, typically by several orders of magnitude.

\begin{figure}[t] \centering
\includegraphics[width=\columnwidth]{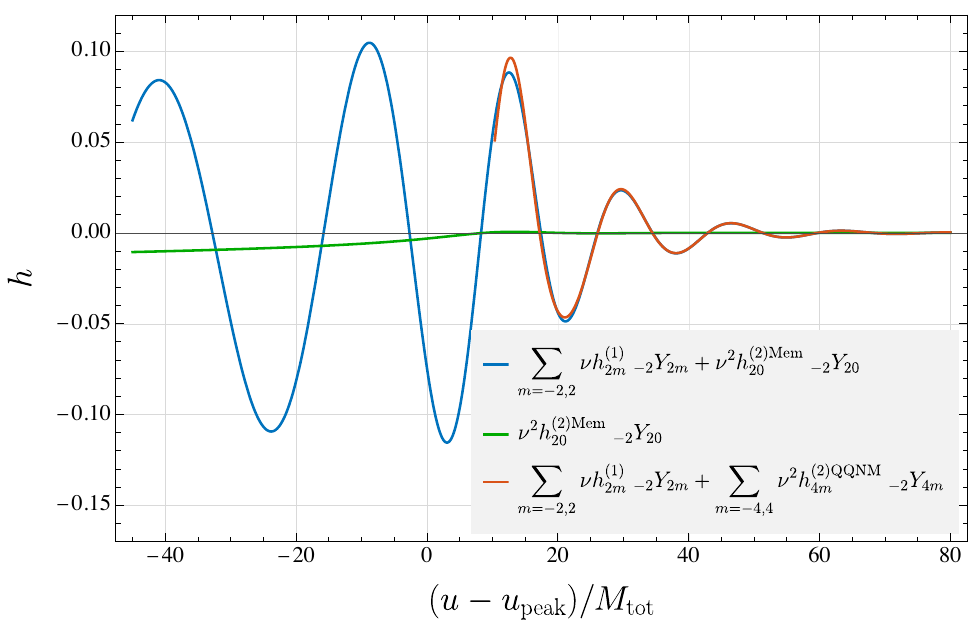}
    \caption{Full first-order strain plus select second-order effects. The blue curve includes the dominant $(\ell,m)=(2,0)$ mode of the nonlinear memory. The red curve, restricted to the green shaded region in Fig.~\ref{fig:Waveform_firstorder_l2m2}, includes the dominant $(\L,\M)=(4,4)$ QQNM mode. The green curve isolates the nonlinear memory's $(\ell,m)=(2,0)$ mode. We have re-expanded in terms of the symmetric mass ratio $\nu\coloneqq\e/(1+\e)^2$ and magnified nonlinearities by considering an equal-mass binary ($\nu=1/4$). The system is viewed edge-on, with $\theta=\pi/2$ and $\phi=0$.} 
    \label{fig:Waveform_secondorder}
\end{figure}

We display the dominant QQNM in the waveform in Figs.~\ref{fig:Waveform_dissection} and~\ref{fig:Waveform_secondorder}. In Fig.~\ref{fig:Waveform_dissection}, we see that the dominant QQNM mode in $h^{(2)}$ is larger than all the overtones in $h^{(1)}$ for times $u\gtrsim u_{\rm peak}+18M$. Although $h^{(2)}$ will be suppressed by a power of $\e$ relative to $h^{(1)}$ in the total waveform, we find the QQNMs will dominate over most linear overtones for moderate mass ratios.

\emph{Memory effects.} Any waveform at $\mathscr{I}^+$, regardless of the spacetime interior, will interact quadratically with itself to generate a new, secular perturbation that leaves test masses permanently displaced~\cite{Christodoulou,Blanchet:1992br,Favata:2010zu,Mitman:2024uss}. We analyzed this effect in detail in small-mass-ratio inspirals in Refs.~\cite{Cunningham:2024dog,Spiers:2026yqx}. 
To explore it here, in the merger-ringdown, we adopt retarded Bondi coordinates $(u,r,\theta^A)$~\cite{Bondi:1962px,Madler:2016xju,Flanagan:2015pxa,Compere:2019gft}, in which the full spacetime metric has components $g_{uu} = -1+2M_{\mathscr{B}}(u,\theta^A)/r + {\cal O}(1/r^2)$ and $g_{AB}=r^2[\Omega_{AB} + r^{-1}C_{AB}(u,\theta^C) + {\cal O}(r^{-2})]$ in the large-$r$ limit, with other components decaying more rapidly and where $\Omega_{AB}$ is the round unit-sphere metric. The Bondi mass aspect $M_{\mathscr{B}}$ encodes the spacetime's mass, linear momentum, and linear supermomentum, and the shear $C_{AB}$ encodes the waveform $h=r^2\bar m^A\bar m^B C_{AB}$~\cite{Flanagan:2015pxa,Compere:2019gft}. The two are related by the exact equation~\cite{Flanagan:2015pxa}
\begin{equation}\label{eq:dmdu}
\dot{M}_{\mathscr{B}} = -\frac{1}{8}\dot C_{AB}\dot C^{AB} + \frac{1}{4}D_A D_B \dot C^{AB},
\end{equation}
where indices are raised with $\Omega^{AB}$, $D_A$ is the covariant derivative compatible with $\Omega_{AB}$, and $\dot{} = \partial/\partial u$. 

At linear order in $\e$, Eq.~\eqref{eq:dmdu} represents a consistency condition, $\dot{M}^{(1)}_{\mathscr{B}} = \frac{1}{4}D_A D_B \dot C^{AB}_{(1)}$. At second order, it involves nonlinear effects:
\begin{align}\label{eq:dm2du}
    \dot{M}_{\mathscr{B}}^{(2)} &= -\frac{1}{8}\dot C^{(1)}_{AB}\dot C^{AB}_{(1)} + \frac{1}{4}D_A D_B \dot C^{AB}_{(2)}.
\end{align}
Following tradition~\cite{Frauendiener:1992dmu,Favata:2010zu,Mitman:2024uss,Cunningham:2024dog}, we divide Eq.~\eqref{eq:dm2du} into linear and nonlinear pieces, $D_A D_B \dot C^{AB}_{(2)\text{Lin}} = 4\dot{M}_{\mathscr{B}}^{(2)}$ and $D_A D_B \dot C^{AB}_{(2)\rm Mem} = \frac{1}{2}\dot C^{(1)}_{AB}\dot C^{AB}_{(1)}$. The nonlinear piece of the shear, $C^{AB}_{(2)\rm Mem}$, is the leading nonlinear memory. Carrying onto third order in Eq.~\eqref{eq:dmdu}, we can identify a nonlinear piece generated by the nonlinear memory's interaction with the first-order waveform, satisfying 
\begin{equation}\label{eq:memory of memory}
D_A D_B \dot C^{AB}_{(3)\text{M-Mem}} = \dot C^{(1)}_{AB}\dot C^{AB}_{(2)\rm Mem}
\end{equation}
---a ``memory of memory''.

Figures~\ref{fig:Waveform_dissection} and \ref{fig:Waveform_secondorder} display the memory and memory of memory in the waveform, computed by solving the above differential equations for $C^{AB}_{(2)\rm Mem}$ and $C^{AB}_{(3)\text{M-Mem}}$, specializing to the remnant BH's super-rest frame~\cite{MaganaZertuche:2021syq,Mitman:2022kwt} in which $C_{AB}(u\to\infty)=0$. The most familiar effect of the memory is the secular offset it creates in the $(2,0)$ mode through the merger regime~\cite{Mitman:2024uss}, shown in green in Fig.~\ref{fig:Waveform_secondorder}. However, in Fig.~\ref{fig:Waveform_dissection} we see both $h^{(2)\rm Mem}$ and $h^{(3)\text{M-Mem}}$ contain a rich structure in the ringdown regime. Such oscillations will be partly degenerate with QQNMs, and one could subtract contributions at QQNM frequencies to isolate the remainder (see broadly related work in~\cite{Shi:2026mrr}). 

Particularly interesting is that, in $C^{AB}_{(3)\text{M-Mem}}$, the nonlinear memory deforms the linear QNMs. To understand this, approximate $C_{AB}^{(1)}$ as a sum of QNMs, in which case  
\begin{equation}\label{eq:mem}
\hspace{-10pt}D_A D_B \dot C^{AB}_{(2)\rm Mem}
 = -\!\!\!\sum_{\bm{n}_1,\bm{n}_2}\!\!\frac{\omega_{\bm{n}_1}\omega_{\bm{n}_2}}{2} C^{(1,\bm{n}_1)}_{AB}C^{AB}_{(1,\bm{n}_2)}e^{-i\omega_{\bm{n}_1\bm{n}_2}u}\!
\end{equation}
with $\omega_{\bm{n}_1\bm{n}_2}\coloneqq(\omega_{\bm{n}_1}+\omega_{\bm{n}_2})$ as in QQNMs. If the two input QNMs in Eq.~\eqref{eq:mem} satisfy ${\rm Re}(\omega_{\bm{n}_1})=-{\rm Re}(\omega_{\bm{n}_2})$, then they generate a non-oscillatory $C^{AB}_{(2)\rm Mem}$. Equation~\eqref{eq:memory of memory} implies this can couple to a QNM to generate a memory of memory
that oscillates with an ordinary QNM frequency ${\rm Re}(\omega_{\bm{n}})$ but with \emph{different} damping rate $-{\rm Im}(\omega_{\bm{n}}+\omega_{\bm{n}_1}+\omega_{\bm{n}_2})$. This is analogous to absorption-induced modes~\cite{Sberna:2021eui, May:2024rrg}, but arising from far-field nonlinearities rather than field dynamics near the BH. It is a small effect in the present case because, as we see in Fig.~\ref{fig:Waveform_secondorder}, there is little overlap between the ringdown and the period when the memory is most secular. But since ${\rm Re}(\omega_{\bm{n}_1})=-{\rm Re}(\omega_{\bm{n}_2})$ requires mirror modes, the effect will be more pronounced in cases with large mirror-mode amplitudes.

\emph{Discussion.} We have performed the first calculations of second-order effects in small-mass-ratio merger-ringdowns. Our framework offers a route to systematically compute nonlinear effects in the merger-ringdown phase of all small-mass-ratio binaries, which can then be combined as building blocks in merger-ringdown models, just as inspiral models combine post-Newtonian, post-Minkowskian, and self-force inputs~\cite{Bini:2019nra,Buonanno:2024byg,Albanesi:2025txj,Honet:2025lmk,Gamboa:2026jht}. Given the high accuracy of second-order inspiral models~\cite{Wardell:2021fyy,Albertini:2022rfe,Mathews:2025txc}, we expect this to be valuable input for merger-ringdown modeling across most mass ratios. The modularity of our approach also allows extensions beyond vacuum GR~\cite{Roy:2025kra}, just as for the inspiral~\cite{Spiers:2023cva,Barsanti:2026ulr,Polcar:2025yto,Rahman:2025mip,Datta:2025ruh,HegadeKR:2025rpr,Vicente:2025gsg,Duque:2025yfm,Zi:2026zpw,Dittmann:2026rtj}, facilitating tests of GR and probes of BHs' environments. 

Our calculations have also illuminated the traditional division of the waveform into an extended inspiral, brief merger, and final ringdown. The high accuracy of an SPA in the extended inspiral shows that almost the whole signal to near the peak amplitude is dominated by direct emission, and the high accuracy of the QNM sum shows that QNMs dominate the ringdown. But the merger regime around the waveform's peak is intransigent: adding more terms to the SPA and to the QNM sum, and adding the branch-cut contribution to the ringdown, does \emph{not} extend the validity of these approximations closer to the peak. Our analysis starkly shows that the traditional QNM description, and the Leaver decomposition it is based on, is not explanatory or useful until times $\gtrsim 10 M$ after the peak, in agreement with when QNM fits to time-domain numerical relativity (NR) waveforms become robust~\cite{Cheung:2023vki,Lim:2022veo,Carullo:2024smg,Mitman:2025hgy}. A valuable followup would be to apply our framework to calculate other nonlinear contributions using more recent decompositions of the linear waveform that \emph{do} extend through the merger~\cite{DeAmicis:2026wqd, Ma:2026hcb, Arnaudo:2025uos, Su:2026fvj}.


\emph{Acknowledgments.} We thank Alex Grant, Badri Krishnan, and Keefe Mitman for helpful discussions about gravitational memory. LK and AP thank Geoffrey Comp\`ere and Loïc Honet for countless discussions about merger-ringdown, Lionel London for inspiring discussions, and Paolo Arnaudo, Gregorio Carullo, Marina De Amicis, Adrien Kuntz, Laura Sberna, and Ben Withers for discussions about dynamical effects in the merger. LK and AP acknowledge the support of the ERC Consolidator/UKRI Frontier Research Grant GWModels (selected by the ERC and funded by UKRI [grant number EP/Y008251/1]). 
KC acknowledges support from the STFC grant number ST/B001170/1.
BL gratefully acknowledges funding from the European Union’s Horizon Europe research and innovation programme under the Marie Sklodowska-Curie grant agreement No. 101209791. 
PB acknowledges support from the Dutch Research Council (NWO) with file number OCENW.M.21.119. 
R.P.M. acknowledges support from the Villum Investigator program supported by the VILLUM Foundation (grant no.\ VIL37766) and the DNRF Chair program (grant no.\ DNRF162) by the Danish National Research Foundation. The Center of Gravity is a Center of Excellence funded by the Danish National Research Foundation under grant No. DNRF184. 
This work makes use of the Black Hole Perturbation Toolkit.

\bibliography{bibliography}

\appendix

\section{End Matter}

\emph{Stationary phase approximation}. The first-order plunge waveform is obtained from Eqs.~\eqref{eq:h1lm} and \eqref{eq:Clm}, which can be written as a double integral over frequency $\omega$ and orbital radius $r'$. Schematically~\cite{Kuchler:2025hwx}, 
\begin{equation}\label{eq:h1SPA}
    h^{(1)}_{\ell m} = \int_{-\infty}^{+\infty} d\omega \int_{2M}^{6M} dr' g_{\ell m}(\omega, r') e^{i \varphi(\omega,r')}
\end{equation}
with $\varphi(\omega, r') \coloneqq \omega \left[t_p(r')-u\right] - m\phi_p(r')$, where $e^{-im\phi_p}$ comes from the point-particle source, $e^{i\omega t_p}$ from the Fourier transform of the source, and $e^{-i\omega u}$ from the inverse Fourier transform. The coordinate time and azimuthal angle along the plunging geodesic 
are given by~\cite{Hadar:2009ip,Folacci:2018cic,Kuchler:2025hwx}
\begin{align}
    t_p(r) &= \frac{2\sqrt{2}(r-24M)}{(6M/r-1)^{1/2}}-44\sqrt{2}M\arctan\!\left[\left(6M/r-1\right)^{1/2}\right]\nonumber\\
    &\quad+4M{\rm arctanh}\left[\frac{1}{\sqrt{2}}(6M/r-1)^{1/2}\right],
    \\[1ex]
    \phi_p(r) &= -\frac{2\sqrt{3}}{(6M/r - 1)^{1/2}}.
\end{align}

In the extended inspiral regime, $g_{\ell m}(\omega, r')$ varies slowly while $e^{i \varphi(\omega,r')}$ oscillates rapidly. The double integral~\eqref{eq:h1SPA} is then dominated by the values of $\omega$ and $r'$ for which $\varphi(\omega,r')$ becomes approximately constant, making it amenable to a two-dimensional SPA~\cite{Kuchler:2025hwx}, which we extend to NLO~\cite{Lewis:2025ydo}:
\begin{equation}\label{eq:SPA}
\begin{split}
    \!h^{(1)\text{SPA}}_{\ell m} =& \left.\frac{2\pi g_{\ell m}(\omega, r')}{|t_p'(r')|}\right\vert_{(\omega, r')=(m\Omega(r_p), r_p)} 
    \\[1ex]
    &+ \frac{i \pi}{|t_p'(r')|^3} \bigg[ 2 t_p'(r')\partial_\omega\partial_{r'} g_{\ell m}(\omega, r')
    \\[1ex]
    &- t_p''(r')\left(2\partial_\omega g_{\ell m}(\omega, r') + \omega\partial^2_\omega g_{\ell m}(\omega, r')\right)
    \\[1ex]
    &+ m \phi_p''(r')\partial^2_\omega g_{\ell m}(\omega, r')\bigg]_{(\omega, r')=(m\Omega(r_p), r_p)},
\end{split}
\end{equation}
where $(\omega, r')=(m\Omega(r_p), r_p)$ is the stationary point for which $(\partial_\omega\varphi,\partial_{r'}\varphi)=(0,0)$.

\begin{figure}[t] \centering
\includegraphics[width=\columnwidth]{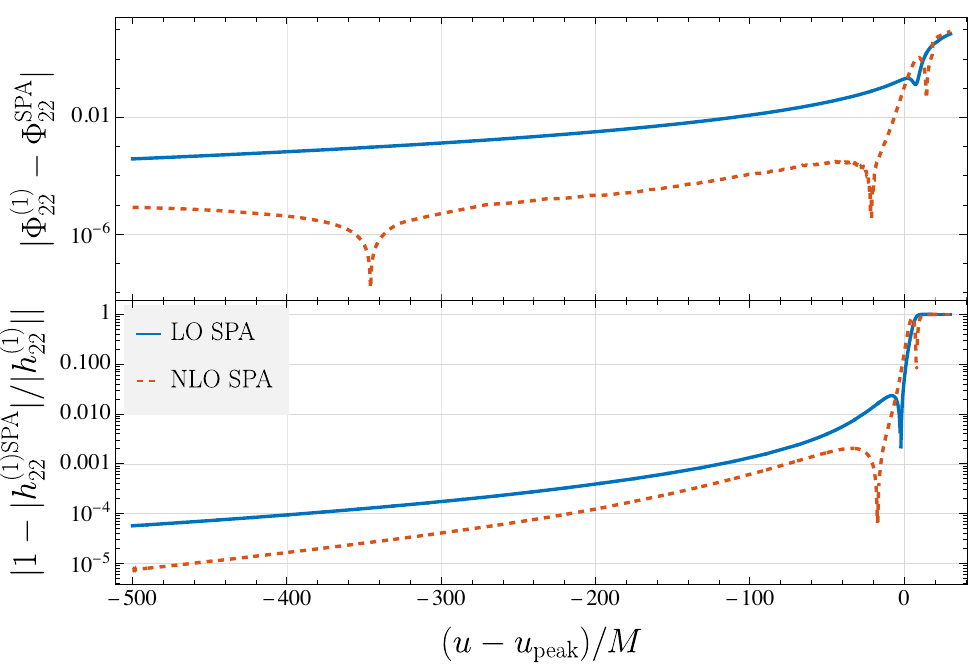}
    \caption{Absolute difference in waveform phase $\Phi^{(1)}_{22}\coloneqq{\rm Arg}(h^{(1)}_{22})$ (top) and relative difference in amplitude $|h^{(1)}_{22}|$ (bottom) between the full waveform and the LO and NLO SPA.} 
    \label{fig:SPA}
\end{figure}

In Fig.~\ref{fig:SPA} we show how including the NLO term in the SPA significantly improves the comparison with the full first-order plunge waveform.

\emph{Nonlinear memory}. Our calculation of nonlinear memory requires a spherical-harmonic decomposition of Eq.~\eqref{eq:dmdu}. We review that decomposition here as well as providing additional numerical results. 

We first expand the Bondi shear in a basis of even- ($Y^{\ell m}_{AB}$) and odd-parity ($X^{\ell m}_{AB}$) tensor harmonics:
\begin{equation}
    C_{AB} = \sum_{\ell=2}^\infty\sum_{m=-\ell}^{+\ell}\left(C^{\ell m}_+ Y^{\ell m}_{AB} + C^{\ell m}_- X^{\ell m}_{AB}\right).
\end{equation}
Using $Y^{\ell m}_{AB} \coloneqq \left[D_A D_B +\frac{1}{2}\ell(\ell+1)\Omega_{AB}\right]Y_{\ell m}$ and $D^AD^BX^{\ell m}_{AB}=0$~\cite{Hopper:2011jlv}, the Ricci identity, and $R^A{}_{BAC}=\Omega_{BC}$, we quickly find
\begin{equation}
D_A D_B \dot C^{AB} = \sum_{\ell m}\frac{D_\ell}{2}\dot C^{\ell m}_+ Y_{\ell m},
\end{equation}
where $Y_{\ell m}$ is an ordinary scalar spherical harmonic and we recall $D_{\ell}\coloneqq(\ell+2)(\ell+1)\ell(\ell-1)$. 

The solution to the nonlinear memory equation, $D_A D_B \dot C^{AB}_{(2)\rm Mem} = \frac{1}{2}\dot C^{(1)}_{AB}\dot C^{AB}_{(1)}$, then reads
\begin{align}\label{eq:C2mem}
    C^{(2)\rm Mem}_{AB} = \sum_{\ell m}\frac{1}{D_{\ell}}\int du\left(\dot C^{(1)}_{CD}\dot C^{CD}_{(1)}\right)_{\ell m}Y^{\ell m}_{AB}.
\end{align}
Here $\left(\dot C^{(1)}_{CD}\dot C^{CD}_{(1)}\right)_{\ell m}:=\int d\Omega\, \left(\dot C^{(1)}_{CD}\dot C^{CD}_{(1)}\right) Y^*_{\ell m}$ can be evaluated in terms of $3j$ symbols; see Eqs.~(9)--(16) of Ref.~\cite{Cunningham:2024dog} or Eq.~(B23) of Ref.~\cite{Bonetto:2021exn}. Analogously, the memory of memory is given by
\begin{equation}\label{eq:C3mem-mem}
    C^{(3)\text{M-Mem}}_{AB} = \sum_{\ell m}\frac{2}{D_{\ell}}\!\int\! du \left(\dot C^{(1)}_{CD}\dot C_{(2)\rm Mem}^{CD}\right)_{\ell m}\!Y^{\ell m}_{AB}.
\end{equation}

\begin{figure}[tb]
\includegraphics[width=\columnwidth]{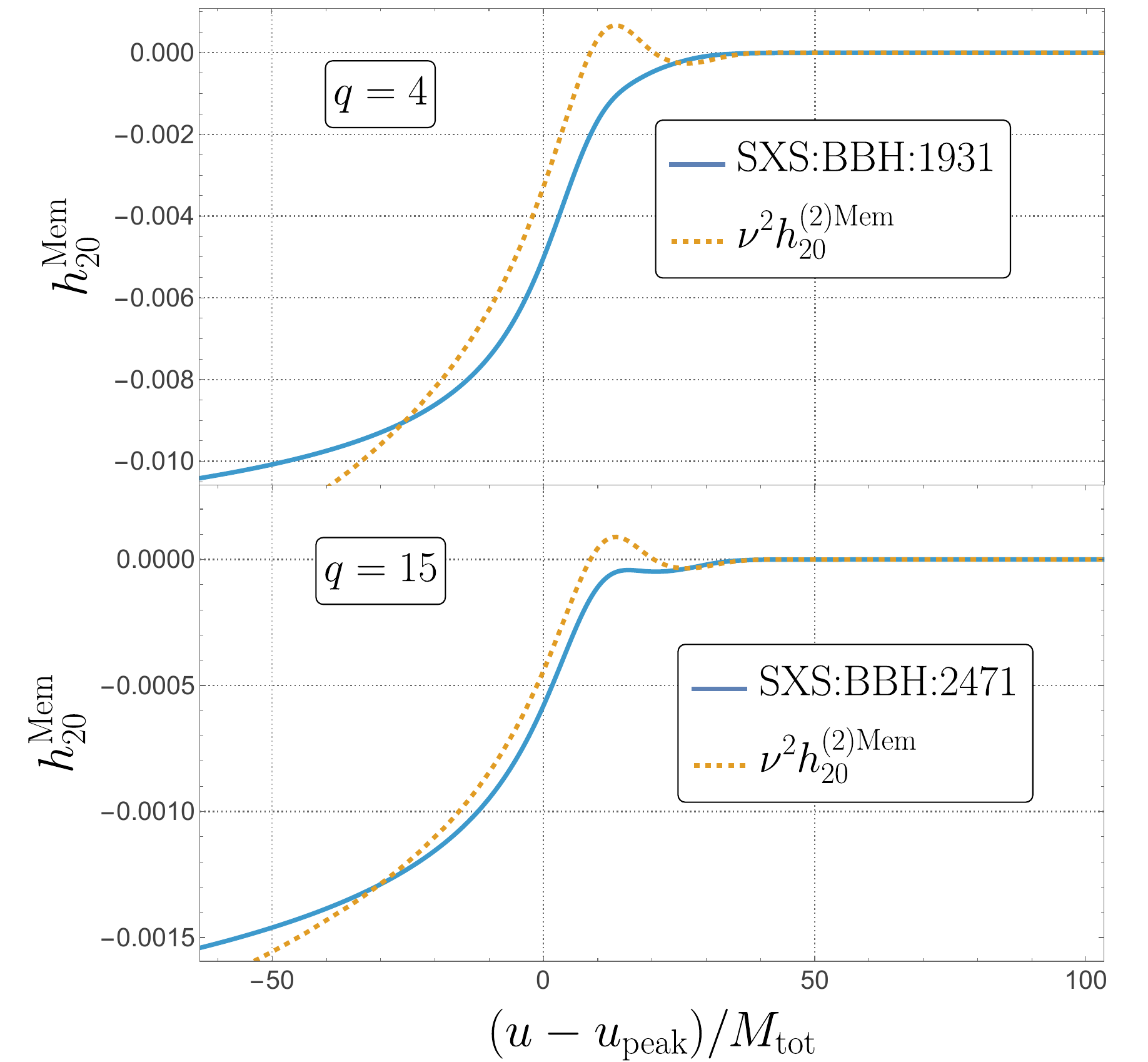} 
\caption{Comparison between the dominant $(\ell,m)=(2,0)$ memory modes calculated from NR (solid blue) and perturbation theory (dashed yellow) for quasicircular binaries with mass ratios $q\coloneqq1/\e=4$ (top) and $q=15$ (bottom). Both NR simulations~\cite{SXS:BBH:1931,SXS:BBH:2471,SXSCatalogData_3.0.0} feature a remnant BH with nearly zero spin. All waveforms are aligned in time at the peaks of their respective $(\ell,m)=(2,2)$ modes.}
\label{fig:NR_Mem_comparison}
\end{figure}

\begin{figure}[t]
    \centering
    \includegraphics[width=\columnwidth]{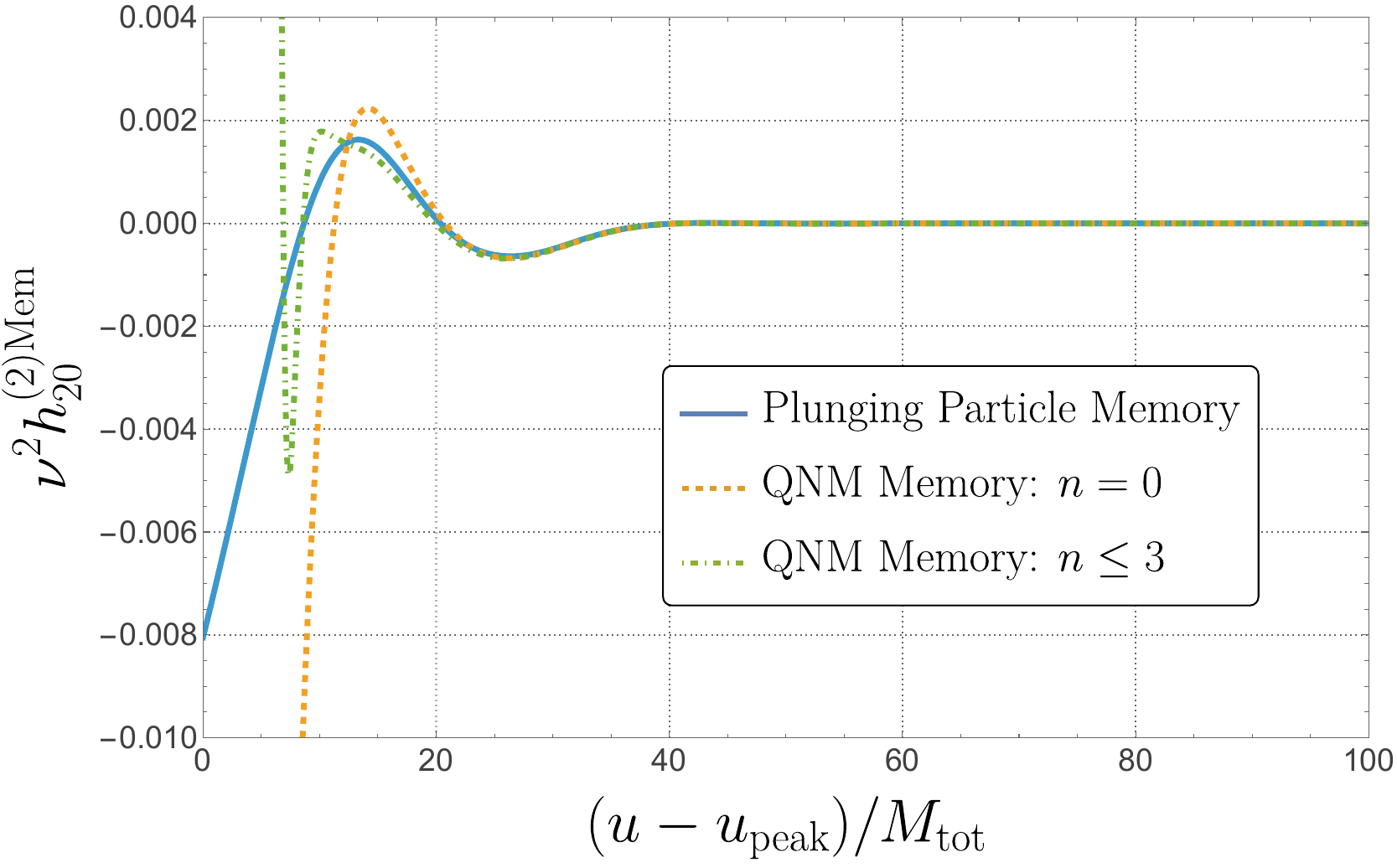}
    \caption{Nonlinear memory in the super-rest frame as calculated from the full plunging-particle waveform (solid blue), from QNMs without overtones (dashed yellow), and from QNMs with overtones up to $n=3$ (dot-dashed green), setting $\nu = 1/4$ for simplicity. As expected from the agreement of the waveforms in Fig.~\ref{fig:Waveform_firstorder_l2m2}, the memory computed from QNMs agrees well at late times with the complete memory, and the two start to disagree near $(u-u_\text{peak})=11 M_{\rm tot}$.}
    \label{fig:QNM_mem}
\end{figure}

Figure~\ref{fig:NR_Mem_comparison} shows a comparison between the memory calculated from perturbation theory ($r^2\bar m^A\bar m^B C^{(2)\text{Mem}}_{AB}$) and from NR simulations, for two mass ratios. In both cases, we select NR simulations that have negligible spin on the final remnant BH, providing the cleanest comparison with our perturbative expansion on a Schwarzschild background. As is standard in such comparisons~\cite{LeTiec:2014oez,Wardell:2021fyy}, we also re-express our perturbative expansion in terms of total mass $M_{\rm tot}$ and symmetric mass ratio $\nu$. We see that the NR and perturbative results converge toward one another for increasing $q$, as expected. 
The main difference is that perturbation theory predicts much larger higher-mode ($\ell>2$) amplitudes during merger. These higher modes produce the characteristic “hump” near $(u-u_\text{peak})\approx 13 M_{\rm tot}$ via coupling between modes with different $\ell$. As $q$ increases, we expect NR amplitudes to approach the perturbative prediction, with relatively more power in higher modes. This trend indeed appears at $q=15$, where the hump is visible in the NR curve albeit smaller.

\begin{figure}[tb]
    \centering
    \includegraphics[width=0.85\columnwidth]{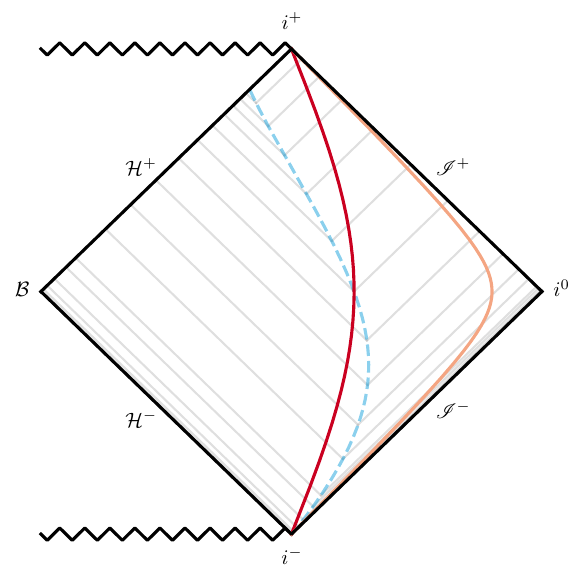}
    \includegraphics[width=0.85\columnwidth]{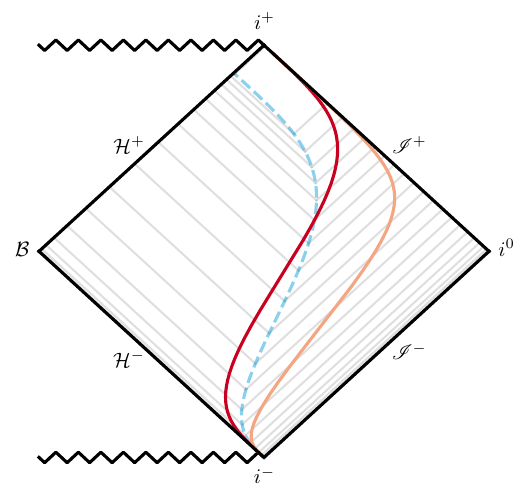}
    \caption{Penrose diagrams with the standard compactification (top) and our modified compactification (bottom). Lines of constant $r$ are highlighted for $r=3M$ (red) and $r=6M$ (orange), along with the plunging geodesic (blue dashed) and lines of constant $u$ or $v$ (gray).}
    \label{fig:penrose_compactification}
\end{figure}

In the special case that $C^{(1)}_{AB}$ is a sum of QNMs, we can write $C^{(1)}_{AB}=\sum_{\bm{n}} C^{(1,\bm{n})}_{AB} e^{-i\omega_{\bm{n}} u}$, where $\bm{n}=(\ell, m, n,\pm)$ includes the spherical-harmonic indices and $C^{(1,\bm{n})}_{AB} := (C^{(1,\bm{n})}_{+}Y^{\ell m}_{AB}+C^{(1,\bm{n})}_{-}X^{\ell m}_{AB})$ includes the tensor harmonics. The integrals and derivatives with respect to $u$ in Eq.~\eqref{eq:C2mem} can then be evaluated to give
\begin{align}
    C^{(2)\rm Mem}_{AB} &= -\sum_{\ell m}\sum_{\bm{n}\bm{n}'}\frac{i\omega_{\bm{n}}\omega_{\bm{n}'}}{D_{\ell}\,\omega_{\bm{n}\bm{n}'}}\nonumber\\
    &\qquad \times\left(C^{(1,\bm{n})}_{CD} C^{CD}_{(1,\bm{n}')}\right)_{\ell m}Y^{\ell m}_{AB} e^{-i\omega_{\bm{n}\bm{n}'}u}
\end{align}
and
\begin{align}
    &C^{(3)\text{M-Mem}}_{AB}  = -2\sum_{\ell m}\sum_{\bm{n}\bm{n}'\bm{n}''}\frac{i\omega_{\bm{n}\bm{n}'}\omega_{\bm{n}''}}{D_{\ell}\,\omega_{\bm{n}\bm{n}'\bm{n}''}}Y^{\ell m}_{AB}\nonumber\\
    &\qquad\qquad \qquad \times\left(C^{(1,\bm{n''})}_{CD} C^{CD}_{(2,\bm{n},\bm{n}')}\right)_{\ell m} e^{-i\omega_{\bm{n}\bm{n}'\bm{n}''}u},
\end{align}
where $\omega_{\bm{n}\bm{n}'}\coloneqq\omega_{\bm{n}}+\omega_{\bm{n}'}$, $\omega_{\bm{n}\bm{n}'\bm{n}''}\coloneqq\omega_{\bm{n}}+\omega_{\bm{n}'}+\omega_{\bm{n}''}$, and
\begin{align}
C_{AB}^{(2,\bm{n},\bm{n}')} &:= -\sum_{\ell m}\frac{i\omega_{\bm{n}}\omega_{\bm{n}'}}{D_{\ell}\,\omega_{\bm{n}\bm{n}'}}\left(C^{(1,\bm{n})}_{CD} C^{CD}_{(1,\bm{n}')}\right)_{\ell m}Y^{\ell m}_{AB}.
\end{align}
Figure~\ref{fig:QNM_mem} compares the memory as computed from QNMs to the memory as computed from our full linear waveform, showing that the distinctive hump in the memory arises from the nonlinear interaction between QNMs.

\emph{Compactification}. Figure~\ref{fig:Penrose_diagram} uses a nonstandard conformal compactification to clearly highlight the SPA and QNM regimes. In the standard compactification, the Kruskal-Szekeres null coordinates, $U = T - R$ and $V = T + R$, are mapped to a finite domain using %
\begin{equation}%
    \label{eq:penrose_mapping}
    x, y = \frac{1}{2} \bigl[ \arctan(V) \pm \arctan(U) \bigr].
\end{equation}
This mapping compresses early and late retarded times, making it difficult to visualise the different regimes in the merger (see Fig.~\ref{fig:penrose_compactification}). 

We remedy this using a nonlinear scaling of the null coordinates,
\begin{equation}
    \tilde{U} = -|T - R|^{-7/2}, \qquad 
    \tilde{V} = \frac{5}{4}(T + R).
\end{equation}
The compact coordinates are then defined as in Eq.~(\ref{eq:penrose_mapping}),  with $(U, V) \rightarrow (\tilde{U},\tilde{V})$. Figure~\ref{fig:penrose_compactification} shows how this transformation expands the spacetime at early and late times.

\end{document}